\documentclass[12pt]{iopart}
\usepackage{graphicx} 
\usepackage{epstopdf} 
\usepackage{epsfig} 
\usepackage{iopams}
\usepackage{setstack}
\usepackage{dcolumn} 
\usepackage{bm} 
\usepackage{braket}
\usepackage{ucs}		
\usepackage[utf8]{inputenc}	
\usepackage[T1]{fontenc}	
\usepackage[english]{babel}	
\usepackage{verbatim}		
\usepackage{color}
\usepackage{indentfirst}		

\usepackage[squaren]{SIunits}
\DeclareRobustCommand{\@NewUnit}[2]{\addunit{#1}{#2}} 
\@NewUnit\gauss{G} 

\usepackage{hyperref}
\hypersetup{
    colorlinks=true,
    linkcolor=blue,
    filecolor=blue,      
    urlcolor=blue,
    citecolor=blue,
    pdftitle={Density matrix reconstruction via Rydberg EIT},
}

\newcommand{\ii}{{\rm i}}
\newcommand{\im}[1]{{\rm Im[#1]}}
\newcommand{\re}[1]{{\rm Re[#1]}}

\begin{document}
\title[Density matrix reconstruction via Rydberg EIT]{Density matrix reconstruction of three-level atoms via Rydberg electromagnetically induced transparency}

\author{V. Gavryusev$^1$, A. Signoles$^1$, M. Ferreira-Cao$^1$, G. Z\"{u}rn$^1$, C.~S.~Hofmann$^1$, G. G\"{u}nter$^1$, H. Schempp$^1$, M.~Robert-de-Saint-Vincent$^1$\footnote{Now at CNRS, Université Paris 13, Laboratoire de Physique des Lasers, Villetaneuse, France.}, S.~Whitlock$^1$ and M.~Weidem\"{u}ller$^{1,2}$}

\address{$^1$Physikalisches Institut, Universit\"{a}t Heidelberg, Im Neuenheimer Feld 226, 69120 Heidelberg, Germany.}
\address{$^2$ Hefei National Laboratory for Physical Sciences at the Microscale and Department of Modern Physics, and CAS Center for Excellence and Synergetic Innovation Center in Quantum Information and Quantum Physics, University of Science and Technology of China, Hefei, Anhui 230026, China. }

\date{\today}

\ead{\mailto{whitlock@physi.uni-heidelberg.de}, \mailto{weidemueller@uni-heidelberg.de}}

\begin{abstract}
We present combined measurements of the spatially-resolved optical spectrum and the total excited-atom number in an ultracold gas of three-level atoms under electromagnetically induced transparency conditions involving high-lying Rydberg states. The observed optical transmission of a weak probe laser at the center of the coupling region exhibits a double peaked spectrum as a function of detuning, whilst the Rydberg atom number shows a comparatively narrow single resonance. By imaging the transmitted light onto a charge-coupled-device camera, we record hundreds of spectra in parallel, which are used to map out the spatial profile of Rabi frequencies of the coupling laser. Using all the information available we can reconstruct the full one-body density matrix of the three-level system, which provides the optical susceptibility and the Rydberg density as a function of spatial position. These results help elucidate the connection between three-level interference phenomena, including the interplay of matter and light degrees of freedom and will facilitate new studies of many-body effects in optically driven Rydberg gases.  
\end{abstract}

\vspace{1pc}
\vspace{1pc}

The experimental and theoretical investigation of ensembles of Rydberg atoms driven by laser fields is currently attracting a great deal of interest~\cite{Comparat2010,Pritchard2013,Loew2012}. For instance, the exceptional properties of Rydberg atoms, such as their tunable long-range interactions and the Rydberg blockade effect, provide new avenues to investigate strongly correlated many-body physics~\cite{Weimer2008,Pohl2010,Bijnen2011,Schwarzkopf2011,Schauss2012,Viteau2012,Hofmann2013,Schempp2014,Malossi2014}, to implement quantum information protocols~\cite{Jaksch2000,Wilk2010,Isenhower2010,Saffman2010,Keating2015,Maller2015} and to create atom-light interfaces operating at the single photon level~\cite{Dudin2012,Peyronel2012,Maxwell2013,Tiarks2014,Gorniaczyk2014}. However, to fully explore the rich physics of these strongly coupled atom-light systems we require new ways to disentangle the different degrees of freedom which link single atom properties to many-body observables. 

Traditionally, in a Rydberg atom experiment, the population of the Rydberg state is measured by field ionization and charged particle detection. While this approach can be sensitive to individual excitations, it provides limited information on the coupled atom-light system and experimentally achieving high spatial resolution is extremely challenging~\cite{Schwarzkopf2011,Schauss2012,Ditzhuijzen2006,Weber2014}. A second approach which is gaining popularity is to map the properties of Rydberg states, e.g. interactions or energy shifts, onto a strong optical transition via an electromagnetically induced transparency (EIT) resonance~\cite{Mohapatra2007,Pritchard2010,Tauschinsky2010,Sen2015}. The transmitted light field can then be readily spatially resolved using a charge-coupled-device (CCD) camera and probed as a function of laser detuning, thereby providing spatially dependent information on the Rydberg state properties~\cite{Tauschinsky2010,Hattermann2012}. However, combining measurements of both the Rydberg population and the transmitted light field in a single experiment potentially allows to probe the underlying system with high spectral and spatial resolution as well as single-atom sensitivity. First steps towards this goal have already been achieved, with the observation of sub-Poissonian statistics of the matter-part of Rydberg dark state polaritons~\cite{Hofmann2013} and electrical readout of Rydberg EIT in thermal vapour cells~\cite{Barredo2013}. Through time resolved measurements of the transparency signal it has been shown that it is possible to reconstruct the Rydberg population~\cite{Mack2015,Karlewski2015}. Additionally, all-optical techniques which exploit the strong interactions between Rydberg "impurity" states and dark-state polaritons enable position and time-resolved measurements of the Rydberg impurity density~\cite{Guenter2012,Guenter2013,Gavryusev2016}.

Here we present simultaneous measurements of the three-level absorption spectrum and Rydberg-state population of an optically driven gas of ultracold atoms. The spectra reveal localized areas with reduced optical absorption of the medium for the probe laser induced by the coupling laser, that we refer to as electromagnetically induced transparency. By combining spatially resolved optical spectroscopy with single-atom sensitive field ionization detection of the Rydberg population in a single experimental setup (Fig.~\ref{fig:setup}\,(a)) we obtain nearly all information on the coupled atom-light system. We describe a procedure to analyse the absorption spectra for hundreds of camera pixels in parallel, each corresponding to a different atomic density and strength of the coupling laser beam. By fitting the data to analytic solutions for the full one-body density matrix in the weak probe regime, we map out the spatially-dependent profile of Rabi coupling frequency which is otherwise hard to determine {\it in-situ}. The spectral width and shape of the Rydberg population resonance measured via field ionization detection is significantly different to that of the transparency resonance measured near the center of the coupling region, which we identify as a consequence of spatial averaging over the excitation volume.

\section{Electromagnetically-induced-transparency and the optical Bloch equations}
\label{sec:EITtheory}

\begin{figure}[!t]
  \centering
  \includegraphics[width=0.7\columnwidth]{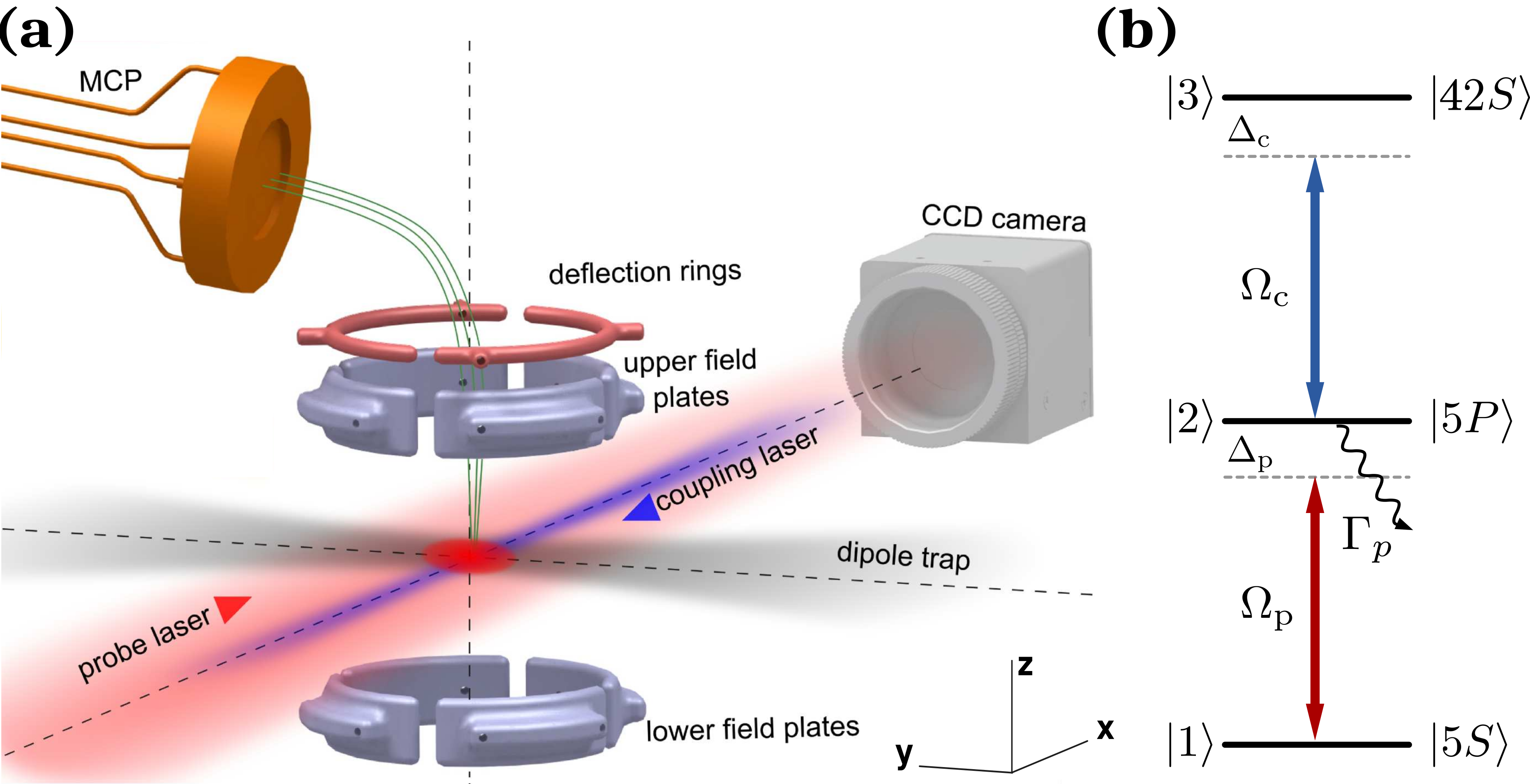}
  \caption{{\bf (a)} Combined optical probing and ionization detection of Rydberg ensembles. $^{87}$Rb atoms are loaded into an optical dipole trap. After trap switch-off the atom cloud expands freely before it is probed under Rydberg-EIT conditions by a probe ($780\,\rm{nm}$) and a coupling ($480\,\rm{nm}$) laser beam. The optical response of the medium to the probe beam is then measured with a CCD camera (imaging system not shown). After image acquisition the Rydberg atoms are field ionized and guided to a micro-channel plate (MCP) detector to measure the Rydberg population. {\bf (b)} Energy level diagram of the three-level system.}
  \label{fig:setup}
\end{figure}

Electromagnetically-induced-transparency is an interference effect which arises in three-level systems composed of two long lived states that are coherently coupled via a short lived state~\cite{Fleischhauer2005}. A strong optical control field renders an otherwise absorbing gas transparent to a weak probe beam, which is accompanied by the evolution of the atomic system into a dark-state that is a superposition of the long-lived states and is decoupled from the light field. We denote the two long-lived states $\vert1\rangle$ and $\vert3\rangle$, which are coupled via a short-lived state $\vert2\rangle$ (see Fig.~\ref{fig:setup}\,(b)). In our experiments, state $\vert3\rangle$ is a high lying Rydberg state and its population can be measured via field ionization detection~\cite{Schempp2010}. Strong laser coupling of the $\vert2\rangle \leftrightarrow \vert3\rangle$ transition with Rabi frequency $\Omega_c$ produces an Autler-Townes doublet of dressed states. Subsequently, the transition amplitudes for the probe laser tuned to the $\vert1\rangle \leftrightarrow \vert2\rangle$ resonance interfere destructively. This gives rise to a transparency resonance which is sensitive to the properties of the Rydberg state.

We start by deriving an approximate analytic solution to the steady state optical Bloch equations (OBE) for the full density matrix $\rho$ for a single atom, which is appropriate in the low density regime. The coupled atom-light system in the rotating wave approximation is described by the following Hamiltonian
\begin{eqnarray}
  \label{eq:Ham}
  \hat H &=& \frac{\hbar}{2}\big(\Omega_p \vert 2 \rangle \langle 1 \vert + \Omega_c \vert 3 \rangle \langle 2 \vert -\Delta_p \vert 2 \rangle \langle 2 \vert - (\Delta_p+ \Delta_c) \vert 3 \rangle \langle 3 \vert + h.c.\big).
\end{eqnarray}
Here $\Omega_{p,c}$ are the Rabi frequencies for the probe and coupling lasers respectively, and $\Delta_{p,c}$ are the detunings from the probe and coupling transitions. In addition, the intermediate state $\vert2\rangle$ spontaneously decays with a rate $\Gamma_p$, whereas decay of the long-lived Rydberg state can be safely neglected. To account for the intermediate state decay and dephasing effects~\cite{Pritchard2010,Gaerttner2013} we express the master equation for the density matrix $\rho$
\begin{equation}
  \label{eq:MastrEq}
  \dot{\rho} = -\frac{i}{\hbar} \left [\hat H,\rho\right ] + \mathcal{L}[\rho] \; ,
\end{equation}
with $\rho_{jk}=\rho_{kj}^{*}$. The Lindblad operator $\mathcal{L}$ includes terms corresponding to the decay of the intermediate state with rate $\Gamma_p$ and to dephasing of the Rydberg state with rate $\gamma_{\rm{deph}}$. This yields the three-level optical Bloch equations
\begin{eqnarray*}
  \label{eq:OBEs}
  \dot{\rho}_{11}&=&-\Omega_p \im{\rho_{12}}+\Gamma_p\rho_{22} \nonumber\\
  \dot{\rho}_{22}&=&\Omega_p \im{\rho_{12}}-\Omega_c
  \im{\rho_{23}}-\Gamma_p\rho_{22} \nonumber\\
  \dot{\rho}_{33}&=&\Omega_c \im{\rho_{23}}\nonumber\\
  \dot{\rho}_{12}&=&-\Gamma_{12}\rho_{12}/2+\ii\Omega_c \rho_{13}/2+\ii\Omega_p(\rho_{11}-\rho_{22})/2 \nonumber\\
  \dot{\rho}_{13}&=&-\Gamma_{13}\rho_{13}/2-\ii(\Omega_p\rho_{23}-\Omega_c \rho_{12})/2 \nonumber\\
  \dot{\rho}_{23}&=&-\Gamma_{23}\rho_{23}/2-\ii(\Omega_c \rho_{33}+\Omega_p\rho_{13}-\Omega_c \rho_{22})/2, \nonumber\\
\end{eqnarray*}
where Im$[x]$ denotes the imaginary part of $x$, and we have defined the rates $\Gamma_{12}=\nobreak\Gamma_p+\nobreak 2\ii\Delta_p$, $\Gamma_{23}=\Gamma_p+\gamma_{\rm{deph}}+2\ii\Delta_c$ and $\Gamma_{13}=\gamma_{\rm{deph}}+2\ii(\Delta_p+\Delta_c)$. By inspection of the OBEs at steady state ($\dot{\rho}=0$), one finds the following relations
\begin{equation}\label{eq:rho33}
  \rho_{22} = \frac{\Omega_p \im{\rho_{12}}}{\Gamma_p}, 
  \hspace{1em} 
  \rho_{23} = \frac{\Omega_p \im{\rho_{13}}}{\re{\Gamma_{23}}},
  \hspace{1em}
  \rho_{33} = \frac{\ii\Gamma_{23}\rho_{23}+\Omega_c\rho_{22}-\Omega_p\rho_{13}}{\Omega_c}.
\end{equation}
Furthermore, in the weak probe limit $(\Omega_p\ll \Omega_c, \Gamma_p)$, we obtain approximate analytic solutions for $\rho_{12}$ and $\rho_{13}$,
\begin{eqnarray}
  \label{eq:rho12}
  \rho_{12}\approx\frac{\ii \Gamma_{13} \Omega_p}{\Gamma_{12}\Gamma_{13}+\Omega_c^2},\hspace{1em} 
  \rho_{13}\approx\frac{-\Omega_c \Omega_p}{\Gamma_{12}\Gamma_{13}+\Omega_c^2},
\end{eqnarray}
which when substituted into equations~(\ref{eq:rho33}) provide solutions for all elements of $\rho$. 

In particular we focus on $\rho_{12}$ and $\rho_{33}$ which relate to observables which can be directly accessed in each experimental run of our system. For this purpose we define the scaled linear optical susceptibility
\begin{equation}
  \label{eq:chi}
  \tilde\chi(\Delta_p)=\frac{\chi(\Delta_p)}{\chi_{\mathrm{2lvl}}(0)}=\frac{\Gamma_p}{\Omega_p}\rho_{12}
\end{equation}
where $\chi_{\mathrm{2lvl}}(0)=6\pi n/k^3$ is the resonant optical susceptibility of a perfect two-level system in the absence of dephasing and power broadening, $n$ is the local atomic density and $k=2\pi/\lambda$ is the wavevector of the probe light field. Provided that $\vert\chi\vert\ll 1$ the transmission of the probe beam is then given by
\begin{equation}
  \label{eq:T}
  T \approx \exp{\biggl(-k\int \im{\chi} dx \biggr)} = \exp{\biggl(-\frac{6\pi}{k^2}\im{\tilde\chi} n_{2d}\biggr)},
\end{equation}
where $n_{2d}$ is the atomic density integrated along the probe propagation direction.

In our experiments, the measured transmission in the regions where the coupling Rabi frequency vanishes ($\Omega_c=0$) provides information on the optical density for the probe transition $6\pi n_{2d}/k^2$ since $\im{\tilde\chi}$ presents a Lorentzian profile $\Gamma_p^2/\vert\Gamma_{12}\vert^2$ as a function of $\Delta_p$. This quantity is then used as a fixed parameter to extract $\im{\tilde\chi}$ for the three-level system including the influence of the coupling beam. Complementary information is provided by field ionization of the gas which provides a direct measure of the integrated Rydberg population $N_r=\nobreak\int\rho_{33}({\bf{r}})n({\bf{r}})d\bf{r}$, where the spatial dependence enters through the atomic density profile $n({\bf{r}})$ and the inhomogeneous profiles of the excitation beams.

\section{Experimental conditions and techniques}
\label{sec:Setup}

General features of our experimental setup are sketched in Fig.~\ref{fig:setup}\,(a) and are detailed in~\cite{Hofmann2014}. We work with $^{87}$Rb atoms with the three-level ladder system: $\vert1\rangle \equiv \vert5S_{1/2},F=2,m_F=2\rangle$, $\vert2\rangle \equiv \vert5P_{3/2},F=3,m_F=3\rangle$, and $\vert3\rangle \equiv \vert42S_{1/2},m_J=1/2\rangle$. A homogeneous magnetic field of $3\,\rm{G}$ is applied along the probe beam direction to define a quantization axis. Approximately $3\cdot10^6$ atoms are loaded into an optical dipole trap at a temperature of $40~\mu$K. We initialize the atoms in the state $\vert1\rangle$ by inducing a microwave transfer via a Landau-Zener sweep of a magnetic field. We control the density of the cloud, without affecting its size, by varying the duration of the sweep to tune the level of adiabaticity of the transfer. Before probing, the trap is switched off and the atoms expand for a fixed time. For an expansion time of $2$~ms the cloud has an ellipsoidal Gaussian shape with $e^{-1/2}$ radii of $\sigma_{\rm{radial}} = 90\pm7\,\rm{\mu m}$, $\sigma_{\rm{axial}} = 380\pm13\,\rm{\mu m}$ and a peak atomic density of $n_0 = 7\cdot10^9\,\rm{cm}^{-3}$. The expected peak Rydberg density is $\rho_{33}n_0\leq 2\cdot10^9\,\rm{cm}^{-3}$. For the $\vert42S_{1/2}\rangle$ state the anticipated blockade radius is $\approx 2.3~\mu$m. This corresponds to a density of $\lesssim 0.1$ Rydberg atoms per blockade volume, therefore the effects of Rydberg-Rydberg interactions can be safely neglected~\cite{Pritchard2010,Schempp2010,Sevincli2011a,Ates2011}. The atomic states are coupled using near resonant laser fields switched on for $t_{exc}=30\,\rm{\mu s}$. The probe laser has a wavelength of $780\,\rm{nm}$, is $\sigma^+$ polarized and is collimated with a Gaussian beam waist of $1.5$~mm. We use a probe Rabi frequency $\Omega_p/2\pi = 1.0\,\rm{MHz}$ which we independently measure using the saturated absorption imaging method \cite{Reinaudi2007}. The atomic cloud is imaged onto a CCD camera with a spatial resolution of $4.8~\mu$m (Rayleigh criterion). A second probe pulse without atoms is used to normalize out the intensity variations and to produce an absorption image. The coupling laser has a wavelength of $480\,\rm{nm}$, is $\sigma^-$ polarized and it is counter aligned to the probe laser. This beam is focused into the center of the cloud with a waist of approximately $15\,\rm{\mu m}$ and an intensity of approximately $0.9\,\rm{kW/cm}^2$. The coupling and probing lasers are both frequency stabilized via the Pound-Drever-Hall method to a high finesse and ultra-stable passive Fabry-P\'{e}rot cavity \cite{Gavryusev2016,Gregory2015} to a linewidth of $\sim 10\,\kilo\hertz$ which is much smaller than the Rydberg-state dephasing rate observed in our experiments.

To measure the optical response we record $93$ absorption images for different probe detunings $\Delta_p/2\pi$ ranging from $-8.1\,\rm{MHz}$ to $8.1\,\rm{MHz}$. We first exclude the pixels illuminated by the coupling beam and perform a fit of each image to a 2D Gaussian distribution reflecting the expected atomic distribution. Interpolating the Gaussian fit into the excluded area allows us to infer the two-level absorption and the local atomic 2D density $n_{2d}$ of the atom cloud at the position of the coupling beam. Using this information and equation~(\ref{eq:T}) we are then able to extract from the measured transmission $T$ the scaled optical susceptibility $\tilde\chi(\Delta_p)$ for each pixel comprising the image of the cloud.

In parallel we also measure the Rydberg-state population by switching on an electric field at the end of the laser pulse to ionize the Rydberg states. Due to the large polarizability of the Rydberg state, the ionizing field causes a sudden shift of the transition frequency by several linewidths in less than $10$~ns, which is fast compared to the criteria for adiabatic following ($\partial \Delta_c/\partial t\ll\Omega_c^{2}/2$), such that the Rydberg population is effectively frozen. The resulting ions are then guided to a microchannel plate detector (MCP) and the detected number of ions is a good measure for the Rydberg population in the excitation volume~\cite{Schempp2010,Hofmann2013}. The guiding field is produced by several electrodes with applied voltages which have been optimized in order to maximize the number of ions reaching the MCP detector~\cite{Hofmann2014}. In these measurements the number of detected ions is large ($>100$) such that overlapping detection events cannot be neglected. Therefore, we integrate the MCP signal over the distribution of arrival times and divide by the average area of a single detection event measured at low excitation number to obtain the number of detected ions. 

\begin{figure}[!t]
  \centering
  \includegraphics[width=0.9\columnwidth]{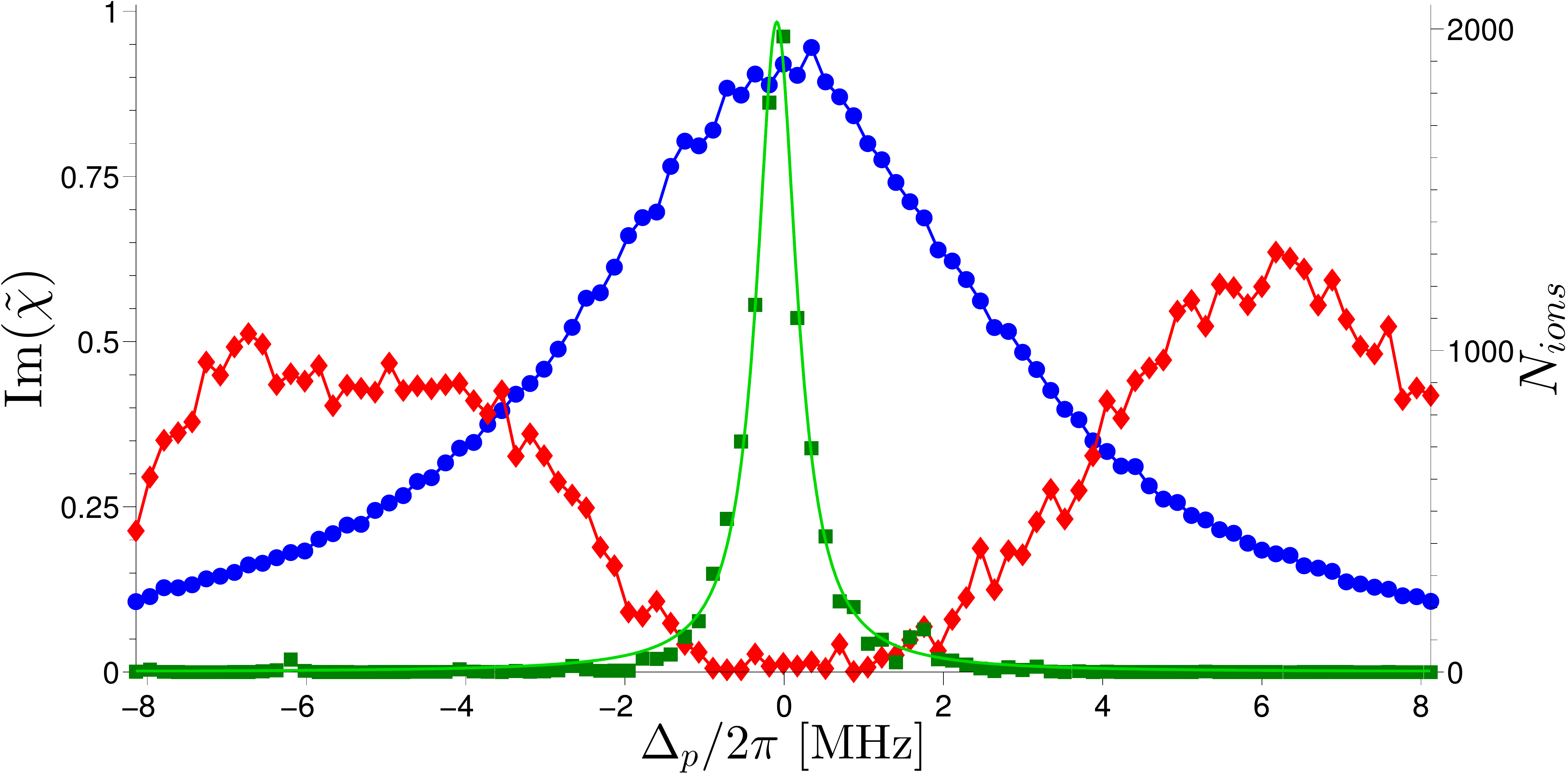}
  \caption{Scaled optical susceptibility and population spectrum as a function of the probe laser detuning. The three measured curves show: three-level optical response averaged over the center of the coupling laser beam (red diamonds), two-level optical response (blue circles) and integrated Rydberg population (green squares - corresponding to the right axis). The green solid line represents the Rydberg population spectrum estimated by numerically solving the OBEs. The slight asymmetry of the absorption spectrum is most likely caused by residual lensing effects due to the non-negligible optical thickness of the sample~\cite{Han2015}.}
  \label{fig:Spectra_comp}
\end{figure}

Figure~\ref{fig:Spectra_comp} shows the imaginary part of the scaled optical susceptibility $\propto \mathrm{Im}[\rho_{12}]$ measured at the center of the coupling laser beam (red diamonds). This is compared with the two-level susceptibility (blue circles) and the Rydberg population spectrum (green squares) measured in the same experimental sequence. On two-photon resonance, and for the parameters given above, we observe an almost complete suppression of absorption ($>99$\%) as compared to the two-level absorption. While the absorption spectrum shows the characteristic double peak shape reflecting the Autler-Townes doublet, the corresponding Rydberg population spectrum is much narrower and shows no evidence for the double peak structure. In the following we will exploit the spatially resolved detection of the transparency feature to elucidate the connection between the transparency resonance and the Rydberg population resonance.

\section{Spatially resolved electromagnetically-induced-transparency}
\label{sec:EITmeas}

\begin{figure}[!t]
  \centering
  \includegraphics[width=0.99\columnwidth]{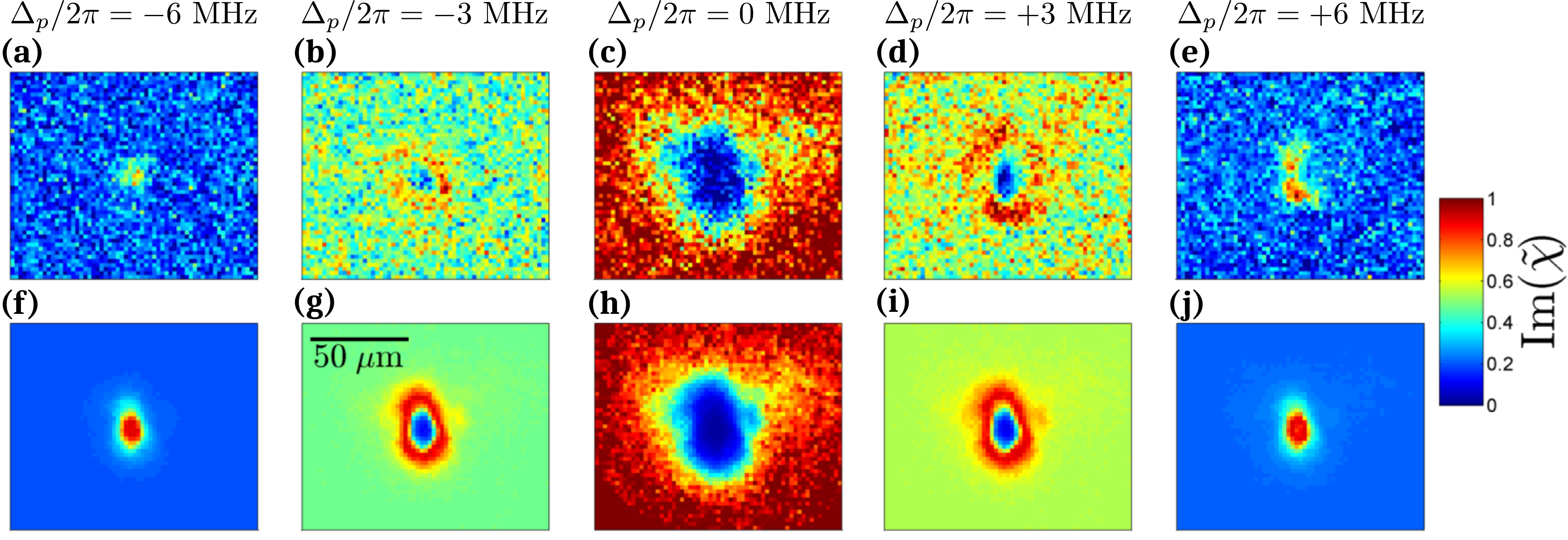}
  \caption{Spatially resolved scaled optical susceptibility $\im{\tilde \chi}$ near the center of the cloud as a function of the probe laser detuning. {\bf(a-e)} Measured cloud response for various probe laser detunings (respectively $-6,-3,0,+3,+6\,\rm{MHz}$) under the same conditions as fig. 2. The region illuminated by the coupling laser is aligned at the center of the region of interest. {\bf(f-j)} Reconstructed spatial response from the OBE steady state solution (from Eqs.~(\ref{eq:rho12}, \ref{eq:chi})), combined with the fitted coupling Rabi frequencies beam profile using the procedure described in the text. The horizontal bar in {\bf(g)} indicates the spatial scale of $50\mu$m.}
  \label{fig:EIT_spatial}
\end{figure}

Figures~\ref{fig:EIT_spatial}\,(a-e) show a selection of extracted scaled optical susceptibilities for different probe detunings, showing the transparency spot at the center of each frame. We find that as a function of detuning the shape and size of the transparency spot varies and exhibits ring-like structures, indicating that the spatial shape of the coupling laser beam plays an important role. To account for this we analyze the optical susceptibility on a pixel-by-pixel basis and construct a series of absorption spectra, one for each position of the cloud. These spectra can then be fit to Eqs.~(\ref{eq:rho12}, \ref{eq:chi}) using the procedure outlined in the following paragraphs which allows for the extraction of the key system parameters in a spatially resolved manner.

A typical dataset involves several thousand pixels which would be prohibitively slow to fit one-by-one. Furthermore the signal-to-noise ratio for a single pixel spectrum is typically quite low. Therefore, we use a two-step fitting algorithm which gives reliable results by minimizing the number of free fit parameters. In the first step we use the two-level peak absorption inferred from the interpolated Gaussian fit to calibrate the two-level optical response including the width of the probe resonance via a fit to a Lorentzian lineshape. For the data shown in Fig.~\ref{fig:EIT_spatial} the resonance width is found to be $(6.21 \pm 0.03)\,\rm{MHz}$ which is in good agreement with the power broadened intermediate state natural decay rate $\Gamma_p/2\pi\cdot\sqrt{1+2(\Omega_p/\Gamma_p)^2} = 6.23\,\rm{MHz}$, confirming that the role of dephasing (e.g. due to laser fluctuations) on this transition is negligible. Analogously, we fit a Lorentzian lineshape to the measured Rydberg population spectrum to determine the coupling laser detuning $\Delta_c/2\pi = (0.10 \pm 0.01)\,\rm{MHz}$ and the width of the resonance $W/2\pi = (0.63 \pm 0.01)\,\rm{MHz}$. This width can be attributed to several effects, but for the coupling Rabi frequency determination we assume it originates entirely from dephasing of the Rydberg state ($\gamma_{\rm{deph}}=W$).

\begin{figure}[!t]
  \centering
  \includegraphics[width=0.99\columnwidth]{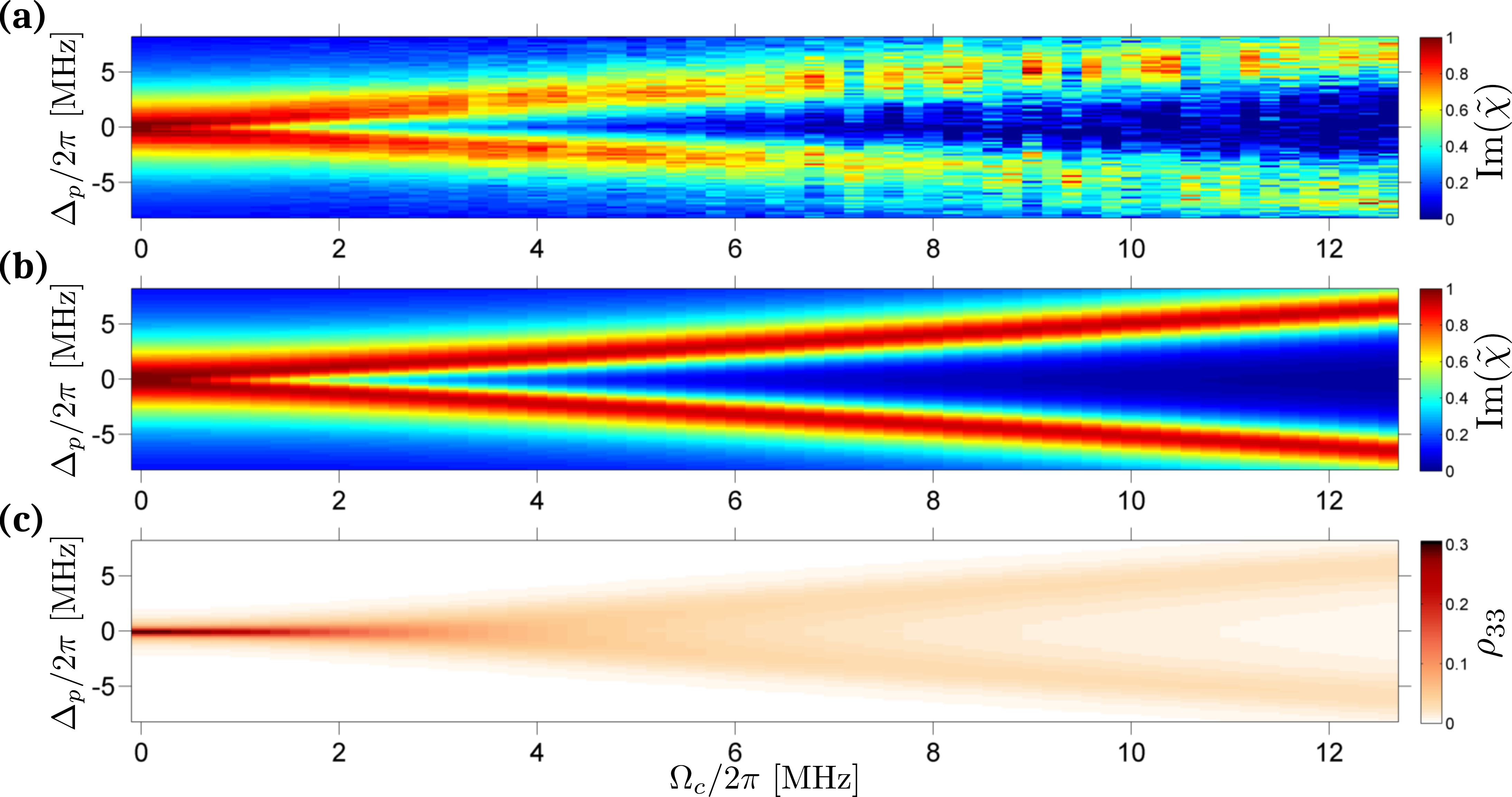}
  \caption{{\bf(a)} Measured three-level absorption spectra (under same conditions as fig. 2) for each pixel, sorted according to the fitted coupling Rabi frequencies with $0.2$ MHz binning. From left to right shows the transition from the EIT regime towards the Autler-Townes regime. {\bf(b)} Fitted absorption spectra using Eqs.~(\ref{eq:rho12}-\ref{eq:chi}) as described in the text. {\bf(c)} Corresponding Rydberg population spectrum calculated analytically using Eq.~(\ref{eq:rho33}).}
  \label{fig:EIT_spectra_fit}
\end{figure}

After the first step all global parameters are fixed and the only remaining free parameter is the value of $\Omega_c$ which differs for each pixel. Using the fitted local two-level resonant optical susceptibility and equations~(\ref{eq:chi}, \ref{eq:T}) we extract the scaled optical susceptibility $\tilde\chi(\Delta_p)$. Fig.~\ref{fig:EIT_spectra_fit}\,(a) shows the measured absorption spectra with each column corresponding to a single pixel in the vicinity of the transparency spot. In order to estimate $\Omega_c(x,y)$ we pre-calculate a set of model absorption spectra corresponding to different values of $\Omega_c/2\pi$ between $0$ and $15$\,MHz in steps of $0.05$\,MHz (Fig.~\ref{fig:EIT_spectra_fit}\,(b)). Minimizing the least-squares difference between each single-pixel spectrum and the model spectra gives an estimate of the local coupling Rabi frequency which best matches the data at each pixel location. The result yields extracted values for $\Omega_c/2\pi$ in the range between $0\,$MHz and $12.8\,$MHz. Compared to the analytic solutions to the OBEs we observe an unexplained slight broadening and reduction of amplitude of the Autler-Townes peaks for large $\Omega_c$. However we find that this discrepancy does not significantly influence the coupling Rabi frequency determination.

\section{\texorpdfstring{Reconstructed spatial distributions of $\rho_{33}$ and $\tilde \chi$}{Reconstructed spatial distributions of rho\_33 and chi}}
\label{sec:DensMatrixRec}

\begin{figure}[!t]
  \centering
  \includegraphics[width=0.8\columnwidth]{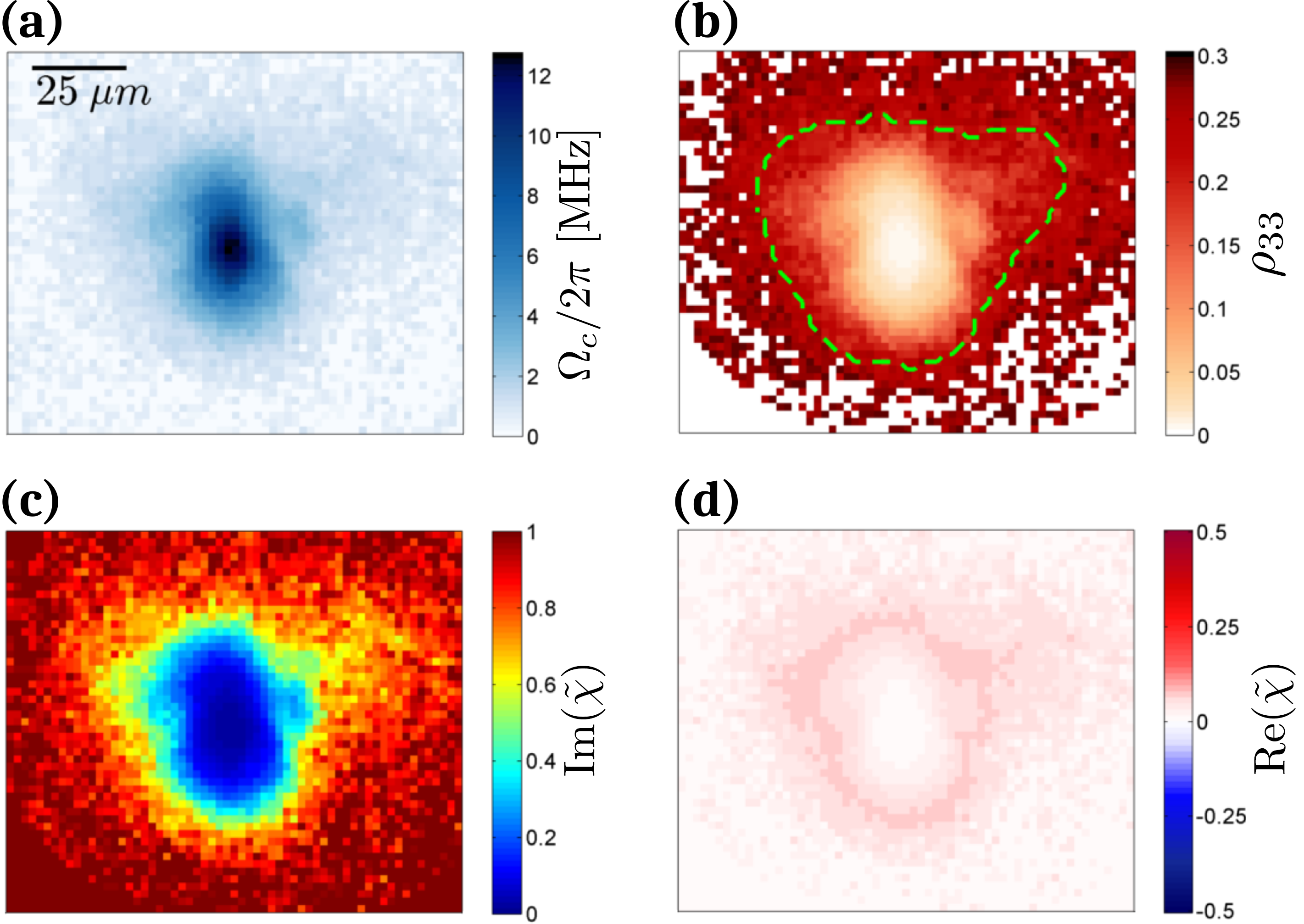}
  \caption{Reconstructed spatial distribution of {\bf(a)} the coupling Rabi frequency $\Omega_c(x,y)$, {\bf(b)} the Rydberg population distribution and {\bf(c-d)} the imaginary and real parts of the scaled optical susceptibility, under the same conditions as the previous figures. For (b-d) we use analytic solutions of the OBE (Eqs.~(\ref{eq:rho33}-\ref{eq:chi})) for $\Delta_p$=0. The green dashed line in (b) marks the limit of validity of the analytic reconstruction of $\rho_{33}$ using Eq.~(\ref{eq:rho33}). Outside of the coupling beam region $\im{\tilde\chi}\rightarrow 1$ (consistent with the two-level response), while $\re{\tilde\chi}$ is approximately zero over the region of interest. The slight increase of $\re{\tilde\chi}$ towards the edges of the coupling beam is due to the relatively steep dispersion for $\Omega_c \ll \Gamma_p$ and the small detuning $\Delta_c/2\pi=0.1$\,MHz.}
  \label{fig:DensMatrixRec}
\end{figure}

Figure~\ref{fig:DensMatrixRec}\,(a) shows the spatial distribution of coupling beam Rabi frequencies as extracted from the fits to the single-pixel optical spectra. The observed shape closely reflects the elliptical shape of the coupling beam which we independently confirm using a beam profiler external to the vacuum system. The measured beam profile is close to a Gaussian with a peak Rabi frequency of $\Omega_c/2\pi=(12.8 \pm 0.5)$~MHz. This is close to the theoretical expectation of $\Omega_c/2\pi=10.7~\rm{MHz}$ taking into account the power of the coupling laser, the dipole matrix element for the $\vert5P_{3/2}\rangle$ to $\vert42S_{1/2}\rangle$ transition, and the spatial extent of the excitation region. The fitted one-sigma radii in the semi-minor and semi-major axes are evaluated to have a size of $11.8\,\rm{\mu m}$ and $15.4\,\rm{\mu m}$ respectively. 

Using the spatially-resolved distribution of Rabi frequencies (Fig.~\ref{fig:DensMatrixRec}\,(a)) and the global parameters constrained by the two-level absorption and the total Rydberg population spectrum, we can reconstruct the full density matrix of the system at each position using the analytical solution derived in equations~(\ref{eq:rho33}) and (\ref{eq:rho12}). As practical examples we show the spatial distribution of the real and imaginary parts of the scaled optical susceptibility $\tilde \chi \propto \rho_{12}$ and the Rydberg population $\rho_{33}$ (Fig.~\ref{fig:DensMatrixRec}\,(b-d)) which demonstrates how both optical and atomic spatially-dependent properties can be reconstructed. 

Using the results of this reconstruction procedure we also show in Fig.~\ref{fig:EIT_spectra_fit}\,(c) the Rydberg population as a function of $\Omega_c$ and $\Delta_p$. For large coupling strengths ($\Omega_c > \Gamma_p$) the Rydberg population resembles the Autler-Townes doublet resonance structure with two maxima at $\Delta_p=\pm \Omega_c/2$, while in the limit of weak coupling $\Omega_c \ll \Gamma_p$ the population is concentrated in a single spectrally narrow resonance with maximum at $\Delta_p=-\Delta_c\approx 0$. The peak population is found for small $\Omega_c$ which can be understood considering that on two photon resonance $\rho_{33}\approx \Omega_p^2/(\Omega_p^2+\Omega_c^2)$ (neglecting dephasing). This is also visible in Fig.~\ref{fig:DensMatrixRec}\,(b) which shows the maximum Rydberg population outside of the coupling beam region. This spatial distribution also explains the spectrally narrow resonance in Fig.~\ref{fig:Spectra_comp} as a consequence of spatially averaging over the entire excitation volume in the field ionization detection. Here we note that at the edges, where $\rho_{33}$ is largest, the validity of the analytic solution for $\rho_{33}$ provided by equations~(\ref{eq:rho33}, \ref{eq:rho12}) is compromised since the condition $\Omega_p\ll\Omega_c$ is not fulfilled. By comparing with numerical solutions of the time-dependent OBEs which are not restricted to the weak probe limit we find that the discrepancy between the approximate and the full numerical solutions remains smaller than $25\%$ for coupling Rabi frequencies $\Omega_c \geq \Omega_p$. This criteria defines an approximate range of validity of the analytic reconstruction, marked by the region inside the green dashed line in Fig.~\ref{fig:DensMatrixRec}\,(b). Furthermore the numerical simulation allows to quantitatively reproduce the measured population spectrum (solid green line in Fig.~\ref{fig:Spectra_comp}) by spatially integrating for each detuning the reconstructed $\rho_{33}(x,y)$. To achieve the best agreement we had to assume a detection efficiency of $\eta=0.4$ and adapt $\gamma_{\rm{deph}}/2\pi$ to the value $(0.20 \pm 0.02)\,\rm{MHz}$. The extra broadening seen in Fig.~\ref{fig:Spectra_comp} can be attributed to power broadening of the $\vert1\rangle \leftrightarrow \vert2\rangle$ transition and spatial averaging over the excitation volume. This dephasing rate is still much larger than the sum of the coupling and probe lasers linewidths, therefore we conclude that it originates from effects associated with atomic motion or residual Rydberg-Rydberg interactions~\cite{Pritchard2010,Gaerttner2013,Zhang2014}. In order to approximately reproduce the measured spectrum using the analytic solution of $\rho_{33}$ (which neglects power broadening), we had to introduce a larger effective dephasing rate $\gamma_{\rm{eff}}/2\pi = 0.37\,\rm{MHz}$ and a cut-off coupling Rabi frequency $\Omega_c^{cut} = 0.35 \Omega_p$, below which we assume the Rydberg population is zero.

The above considerations only weakly influence the determination of the coupling Rabi frequency $\Omega_c$ since dephasing has a minimal effect on the Autler-Townes splitting. We also verify that it has a small effect on the reconstructed components of the density matrix and optical susceptibility in the considered range. The reconstructed imaginary and real parts of the scaled optical susceptibility $\tilde\chi$ using the analytical formulas in Sec.~\ref{sec:EITtheory} are shown for $\Delta_p=0$ (for $\gamma_{\rm{deph}}/2\pi=0.63\,\rm{MHz}$) in Fig.~\ref{fig:DensMatrixRec}\,(c-d). $\im{\tilde\chi}$ relates to the absorption coefficient of light propagating through the medium. As expected, it shows almost full transparency $\im{\tilde\chi}\approx 0$ at the center of the coupling beam, whereas it approaches the two-level response outside of the coupling beam region. The reconstructed $\im{\tilde\chi}$ is also plotted for different detunings in figures~\ref{fig:EIT_spatial}\,(f-j), showing good qualitative agreement with the experimental measurements. However we note a slight asymmetry in the detuning dependence of the experimental data which could be attributed to lensing effects which are not accounted for in our simple model. In contrast, $\re{\tilde\chi}$, which is responsible for light dispersion, is nearly zero across the whole spatial profile for $\Delta_p=0$. The small observed deviation at the edges of the coupling beam region is due to the relatively steep dispersion for $\Omega_c \ll \Gamma_p$, combined with the slight detuning of the coupling beam $\Delta_c/2\pi=0.1$\,MHz. For larger detunings $\Delta_p\approx \Omega_c/2$ the amplitude of $\re{\tilde\chi}$ can increase significantly which could be responsible for the lensing effects seen in Figs.~\ref{fig:EIT_spatial}\,(a-e). Analogous effects have been recently studied in Ref.~\cite{Han2015} and will be the topic of future work.

\section{Conclusion}

By combining field ionization detection with optical spectroscopy under Rydberg EIT conditions, we have reconstructed the full single-atom density matrix of the system thereby obtaining nearly full information about the coupled atom-light system. Spatially resolving the absorption spectra and analyzing hundreds of camera pixels in parallel gives information on hundreds of mesoscopic Rydberg ensembles, each with different densities or laser parameters. The extracted spatially-dependent profile of Rabi coupling frequencies explains the observed spectral shape and width of the Rydberg population resonance as a consequence of spatially averaging over the entire excitation volume in the field ionization detection.

The combination of optical and population-based probing of coherently driven three-level atomic systems as realised in these experiments offers new avenues for studying multilevel interference effects such as electromagnetically-induced-transparency, coherent population trapping and stimulated-Raman adiabatic passage with simultaneous access to all degrees of freedom. Furthermore, the reconstructed spatially-dependent Rabi frequency, Rydberg population and optical susceptibility serve as valuable input for modeling light propagation in interacting Rydberg ensembles~\cite{Ates2011,Gorshkov2013,GaerttnerPRA2014,Bienias2014} and realizing new non-destructive imaging techniques for strongly-interacting particles, with single atom sensitivity~\cite{Guenter2012,Guenter2013,Gavryusev2016,Olmos2011}. Ultimately, these efforts complemented by the technique described here, will enable new studies of the correlations between atoms and photons induced by Rydberg-Rydberg interactions, relevant for example to current and future studies of nonlinear light propagation in strongly interacting media~\cite{Peyronel2012,Sevincli2011,Stanojevic2013,Tresp2015,Moos2015} and Rydberg dressed quantum fluids~\cite{Helmrich2016,Gaul2016,DeSalvo2016} which exploit strong-atom light coupling in three-level atomic systems.

\section*{Acknowledgments}
We thank M. G\"{a}rttner, J. Evers and M. Fleischhauer for fruitful discussions. This work is supported in part by the Heidelberg Center for Quantum Dynamics and by the Deutsche Forschungsgemeinschaft under WE2661/10.2. C.S.H. and G.G. acknowledge support by the Studienstiftung des deutschen Volkes, C.S.H. from the Landesgraduierten Akademie, S.W. from the Deutsche Forschungsgemeinschaft (under WH141/1.1). M.R.D.S.V. (grant No. FP7-PEOPLE-2011-IEF-300870), M.F.C. and V.G. (grant No. FP7-PEOPLE-2010-ITN-265031) acknowledge support from the EU Marie-Curie program, and V.G. by the IMPRS-QD.

\section*{References}
\bibliographystyle{iopart-num}
\bibliography{paper_comb_det}

\providecommand{\newblock}{}
\begin{thebibliography}{10}
\expandafter\ifx\csname url\endcsname\relax
  \def\url#1{{\tt #1}}\fi
\expandafter\ifx\csname urlprefix\endcsname\relax\def\urlprefix{URL }\fi
\providecommand{\eprint}[2][]{\url{#2}}

\bibitem{Comparat2010}
Comparat D and Pillet P 2010 {\em J. Opt. Soc. Am. B\/} {\bf 27} A208

\bibitem{Pritchard2013}
Pritchard J~D, Weatherill K~J and Adams C~S 2013 {\em Annual Review of Cold
  Atoms and Molecules\/} vol~1 (World Scientific) chap 8 - Nonlinear optics
  using cold Rydberg atoms, pp 301--350

\bibitem{Loew2012}
L\"{o}w R, Weimer H, Nipper J, Balewski J~B, Butscher B, B\"{u}chler H~P and
  Pfau T 2012 {\em Journal of Physics B: Atomic, Molecular and Optical
  Physics\/} {\bf 45} 113001

\bibitem{Weimer2008}
Weimer H, L{\"o}w R, Pfau T and B{\"u}chler H~P 2008 {\em Phys. Rev. Lett.\/}
  {\bf 101} 250601

\bibitem{Pohl2010}
Pohl T, Demler E and Lukin M~D 2010 {\em Phys. Rev. Lett.\/} {\bf 104} 043002

\bibitem{Bijnen2011}
van Bijnen R~M~W, Smit S, van Leeuwen K~A~H, Vredenbregt E~J~D and Kokkelmans
  S~J~J~M~F 2011 {\em Journal of Physics B: Atomic, Molecular and Optical
  Physics\/} {\bf 44} 184008

\bibitem{Schwarzkopf2011}
Schwarzkopf A, Sapiro R~E and Raithel G 2011 {\em Phys. Rev. Lett.\/} {\bf 107}
  103001

\bibitem{Schauss2012}
Schau\ss{} P, Cheneau M, Endres M, Fukuhara T, Hild S, Omran A, Pohl T, Gross
  C, Kuhr S and Bloch I 2012 {\em Nature\/} {\bf 491} 87--91

\bibitem{Viteau2012}
Viteau M, Huillery P, Bason M~G, Malossi N, Ciampini D, Morsch O, Arimondo E,
  Comparat D and Pillet P 2012 {\em Phys. Rev. Lett.\/} {\bf 109} 053002

\bibitem{Hofmann2013}
Hofmann C~S, G\"{u}nter G, Schempp H, Robert-de Saint-Vincent M, G\"{a}rttner
  M, Evers J, Whitlock S and Weidem\"{u}ller M 2013 {\em Phys. Rev. Lett.\/}
  {\bf 110} 203601

\bibitem{Schempp2014}
Schempp H, G\"unter G, Robert-de Saint-Vincent M, Hofmann C~S, Breyel D, Komnik
  A, Sch\"onleber D~W, G\"arttner M, Evers J, Whitlock S and Weidem\"uller M
  2014 {\em Phys. Rev. Lett.\/} {\bf 112} 013002

\bibitem{Malossi2014}
Malossi N, Valado M~M, Scotto S, Huillery P, Pillet P, Ciampini D, Arimondo E
  and Morsch O 2014 {\em Phys. Rev. Lett.\/} {\bf 113} 023006

\bibitem{Jaksch2000}
Jaksch D, Cirac J~I, Zoller P, Rolston S~L, C\^ot\'e R and Lukin M~D 2000 {\em
  Phys. Rev. Lett.\/} {\bf 85} 2208--2211

\bibitem{Wilk2010}
Wilk T, Ga\"etan A, Evellin C, Wolters J, Miroshnychenko Y, Grangier P and
  Browaeys A 2010 {\em Phys. Rev. Lett.\/} {\bf 104} 010502

\bibitem{Isenhower2010}
Isenhower L, Urban E, Zhang X~L, Gill A~T, Henage T, Johnson T~A, Walker T~G
  and Saffman M 2010 {\em Phys. Rev. Lett.\/} {\bf 104} 010503

\bibitem{Saffman2010}
Saffman M, Walker T~G and M{\o}lmer K 2010 {\em Reviews of Modern Physics\/}
  {\bf 82} 2313--2363

\bibitem{Keating2015}
Keating T, Cook R~L, Hankin A~M, Jau Y~Y, Biedermann G~W and Deutsch I~H 2015
  {\em Phys. Rev. A\/} {\bf 91} 012337

\bibitem{Maller2015}
Maller K~M, Lichtman M~T, Xia T, Sun Y, Piotrowicz M~J, Carr A~W, Isenhower L
  and Saffman M 2015 {\em Phys. Rev. A\/} {\bf 92} 022336

\bibitem{Dudin2012}
Dudin Y~O and Kuzmich A 2012 {\em Science\/} {\bf 336} 887 -- 889

\bibitem{Peyronel2012}
Peyronel T, Firstenberg O, Liang Q~Y, Hofferberth S, Gorshkov A~V, Pohl T,
  Lukin M~D and Vuleti\'{c} V 2012 {\em Nature\/} {\bf 488} 57--60

\bibitem{Maxwell2013}
Maxwell D, Szwer D~J, Paredes-Barato D, Busche H, Pritchard J~D, Gauguet A,
  Weatherill K~J, Jones M~P~A and Adams C~S 2013 {\em Phys. Rev. Lett.\/} {\bf
  110} 103001

\bibitem{Tiarks2014}
Tiarks D, Baur S, Schneider K, D\"{u}rr S and Rempe G 2014 {\em Phys. Rev.
  Lett.\/} {\bf 113} 053602

\bibitem{Gorniaczyk2014}
Gorniaczyk H, Tresp C, Schmidt J, Fedder H and Hofferberth S 2014 {\em Phys.
  Rev. Lett.\/} {\bf 113} 053601

\bibitem{Ditzhuijzen2006}
van Ditzhuijzen C~S~E, Koenderink A~F, Noordam L~D and van Linden van~den
  Heuvell H~B 2006 {\em Eur. Phys. J. D\/} {\bf 40} 13

\bibitem{Weber2014}
Weber T~M, Honing M, Niederprum T, Manthey T, Thomas O, Guarrera V,
  Fleischhauer M, Barontini G and Ott H 2015 {\em Nature Physics\/} {\bf 11}
  157--161

\bibitem{Mohapatra2007}
Mohapatra A~K, Jackson T~R and Adams C~S 2007 {\em Phys. Rev. Lett.\/} {\bf 98}
  113003

\bibitem{Pritchard2010}
Pritchard J~D, Maxwell D, Gauguet A, Weatherill K~J, Jones M~P~A and Adams C~S
  2010 {\em Phys. Rev. Lett.\/} {\bf 105} 193603

\bibitem{Tauschinsky2010}
Tauschinsky A, Thijssen R~M~T, Whitlock S, van Linden van~den Heuvell H~B and
  Spreeuw R~J~C 2010 {\em Phys. Rev. A\/} {\bf 81} 063411

\bibitem{Sen2015}
Sen S, Dey T~K, Nath M~R and Gangopadhyay G 2015 {\em Journal of Modern
  Optics\/} {\bf 62} 166--174

\bibitem{Hattermann2012}
Hattermann H, Mack M, Karlewski F, Jessen F, Cano D and Fort\'agh J 2012 {\em
  Phys. Rev. A\/} {\bf 86} 022511

\bibitem{Barredo2013}
Barredo D, K\"ubler H, Daschner R, L\"ow R and Pfau T 2013 {\em Phys. Rev.
  Lett.\/} {\bf 110} 123002

\bibitem{Mack2015}
Mack M, Grimmel J, Karlewski F, S\'ark\'any L~m~H, Hattermann H and Fort\'agh J
  2015 {\em Phys. Rev. A\/} {\bf 92} 012517

\bibitem{Karlewski2015}
Karlewski F, Mack M, Grimmel J, S\'andor N and Fort\'agh J 2015 {\em Phys. Rev.
  A\/} {\bf 91} 043422

\bibitem{Guenter2012}
G\"{u}nter G, Robert-de Saint-Vincent M, Schempp H, Hofmann C~S, Whitlock S and
  Weidem\"{u}ller M 2012 {\em Phys. Rev. Lett.\/} {\bf 108} 013002

\bibitem{Guenter2013}
G\"{u}nter G, Schempp H, Robert-de Saint-Vincent M, Gavryusev V, Helmrich S,
  Hofmann C~S, Whitlock S and Weidem\"{u}ller M 2013 {\em Science\/} {\bf 342}
  954--956

\bibitem{Gavryusev2016}
Gavryusev V, Ferreira-Cao M, Keki{\'c} A, Z{\"u}rn G and Signoles A 2016 {\em
  ArXiv e-prints\/} (\textit{Preprint} \eprint{1602.04143})

\bibitem{Fleischhauer2005}
Fleischhauer M, Imamoglu A and Marangos J~P 2005 {\em Reviews of Modern
  Physics\/} {\bf 77} 633--673

\bibitem{Schempp2010}
Schempp H, G\"unter G, Hofmann C~S, Giese C, Saliba S~D, DePaola B~D, Amthor T,
  Weidem\"uller M, Sevin\c{c}li S and Pohl T 2010 {\em Phys. Rev. Lett.\/} {\bf
  104} 173602

\bibitem{Gaerttner2013}
G\"{a}rttner M and Evers J 2013 {\em Phys. Rev. A\/} {\bf 88}(3) 033417

\bibitem{Hofmann2014}
Hofmann C, G\"{u}nter G, Schempp H, M\"{u}ller N, Faber A, Busche H, Robert-de
  Saint-Vincent M, Whitlock S and Weidem\"{u}ller M 2014 {\em Frontiers of
  Physics\/} {\bf 9} 571--586

\bibitem{Sevincli2011a}
Sevin\c{c}li S, Ates C, Pohl T, Schempp H, Hofmann C~S, G\"{u}nter G, Amthor T,
  Weidem\"{u}ller M, Pritchard J~D, Maxwell D, Gauguet A, Weatherill K~J, Jones
  M~P~A and Adams C~S 2011 {\em Journal of Physics B: Atomic, Molecular and
  Optical Physics\/} {\bf 44} 184018

\bibitem{Ates2011}
Ates C, Sevin\c{c}li S and Pohl T 2011 {\em Phys. Rev. A\/} {\bf 83} 041802

\bibitem{Reinaudi2007}
Reinaudi G, Lahaye T, Wang Z and Gu\'{e}ry-Odelin D 2007 {\em Opt. Lett.\/}
  {\bf 32} 3143--3145

\bibitem{Gregory2015}
Gregory P~D, Molony P~K, Köppinger M~P, Kumar A, Ji Z, Lu B, Marchant A~L and
  Cornish S~L 2015 {\em New Journal of Physics\/} {\bf 17} 055006

\bibitem{Han2015}
Han J, Vogt T, Manjappa M, Guo R, Kiffner M and Li W 2015 {\em Phys. Rev. A\/}
  {\bf 92}(6) 063824

\bibitem{Zhang2014}
Zhang H, Zhang L, Wang L, Bao S, Zhao J, Jia S and Raithel G 2014 {\em Phys.
  Rev. A\/} {\bf 90}(4) 043849

\bibitem{Gorshkov2013}
Gorshkov A~V, Nath R and Pohl T 2013 {\em Phys. Rev. Lett.\/} {\bf 110}(15)
  153601

\bibitem{GaerttnerPRA2014}
G\"{a}rttner M, Whitlock S, Sch\"{o}nleber D~W and Evers J 2014 {\em Phys. Rev.
  A\/} {\bf 89} 063407

\bibitem{Bienias2014}
Bienias P, Choi S, Firstenberg O, Maghrebi M~F, Gullans M, Lukin M~D, Gorshkov
  A~V and B\"{u}chler H~P 2014 {\em Phys. Rev. A\/} {\bf 90} 053804

\bibitem{Olmos2011}
Olmos B, Li W, Hofferberth S and Lesanovsky I 2011 {\em Phys. Rev. A\/} {\bf
  84} 041607

\bibitem{Sevincli2011}
Sevin\c{c}li S, Henkel N, Ates C and Pohl T 2011 {\em Phys. Rev. Lett.\/} {\bf
  107} 153001

\bibitem{Stanojevic2013}
Stanojevic J, Parigi V, Bimbard E, Ourjoumtsev A and Grangier P 2013 {\em Phys.
  Rev. A\/} {\bf 88} 053845

\bibitem{Tresp2015}
Tresp C, Bienias P, Weber S, Gorniaczyk H, Mirgorodskiy I, B{\"{u}}chler H~P
  and Hofferberth S 2015 {\em Phys. Rev. Lett.\/} {\bf 115} 083602

\bibitem{Moos2015}
Moos M, H\"{o}ning M, Unanyan R and Fleischhauer M 2015 {\em Phys. Rev. A\/}
  {\bf 92} 053846

\bibitem{Helmrich2016}
Helmrich S, Arias A, Pehoviak N and Whitlock S 2016 {\em Journal of Physics B:
  Atomic, Molecular and Optical Physics\/} {\bf 49} 03LT02

\bibitem{Gaul2016}
Gaul C, DeSalvo B~J, Aman J~A, Dunning F~B, Killian T~C and Pohl T 2016 {\em
  Phys. Rev. Lett.\/} {\bf 116} 243001

\bibitem{DeSalvo2016}
DeSalvo B~J, Aman J~A, Gaul C, Pohl T, Yoshida S, Burgd\"orfer J, Hazzard
  K~R~A, Dunning F~B and Killian T~C 2016 {\em Phys. Rev. A\/} {\bf 93} 022709

\end{thebibliography}

\end{document}